\documentclass[aps, prl, superscriptaddress, twocolumn]{revtex4-1}
\usepackage[utf8]{inputenc}
\usepackage[margin=0.6in]{geometry}
\usepackage{graphicx}
\usepackage[table,xcdraw]{xcolor}
\usepackage{amsmath,amssymb,mathtools}
\usepackage{booktabs}
\usepackage{multirow}
\usepackage{braket}
\usepackage{blindtext}
\usepackage{float}
\usepackage{gensymb}
\usepackage{subcaption}
\usepackage[export]{adjustbox}

\usepackage[toc,page]{appendix}
\usepackage[colorlinks=true,pdfborder={0 0 0},linkcolor=blue, citecolor=blue, allcolors = blue]{hyperref}
\graphicspath{{Images/}}

\begin{document}

\renewcommand{\abstractname}{\vspace{-\baselineskip}}

\title{Tensor gradiometry with a diamond magnetometer}

\author{A. J. Newman}
\author{S. M. Graham}
\affiliation{\footnotesize Department of Physics, University of Warwick, Gibbet Hill Road, Coventry, CV4 7AL, United Kingdom}
\affiliation{\footnotesize EPSRC Centre for Doctoral Training in Diamond Science and Technology, University of Warwick, Coventry CV4 7AL, United Kingdom}

\author{A. M. Edmonds}
\author{D. J. Twitchen}
\author{M. L. Markham}
\affiliation{\footnotesize Element Six Innovation, Fermi Avenue, Harwell Oxford, Didcot OX11 0QR Oxfordshire, United Kingdom}

\author{G. W. Morley}
\affiliation{\footnotesize Department of Physics, University of Warwick, Gibbet Hill Road, Coventry, CV4 7AL, United Kingdom}
\affiliation{\footnotesize EPSRC Centre for Doctoral Training in Diamond Science and Technology, University of Warwick, Coventry CV4 7AL, United Kingdom}

\begin{abstract}

Vector magnetometry provides more information than scalar measurements for magnetic surveys utilized in space, defense, medical, geological and industrial applications. These areas would benefit from a mobile vector magnetometer that can operate in extreme conditions. Here we present a scanning fiber-coupled nitrogen vacancy (NV) center vector magnetometer. Feedback control of the microwave excitation frequency is employed to improve dynamic range and maintain sensitivity during movement of the sensor head. Tracking of the excitation frequency shifts for all four orientations of the NV center allow us to image the vector magnetic field of a damaged steel plate. We calculate the magnetic tensor gradiometry images in real time, and they allow us to detect smaller damage than is possible with vector or scalar imaging.  

\end{abstract}

\maketitle

Tensor gradiometry calculates the gradient of all three components of a vector in all three directions \cite{Schiffler_Tensor_Grad, luheshi_2021, KEBEDE2021e06872}. At each point in space, magnetic tensor gradiometry measures 

\begin{equation}
    \begin{bmatrix}
        \mathrm{B_{xx}} & \mathrm{B_{yx}} & \mathrm{B_{zx}} \\ 
        \mathrm{B_{xy}} & \mathrm{B_{yy}} & \mathrm{B_{zy}} \\
        \mathrm{B_{xz}} & \mathrm{B_{yz}} & \mathrm{B_{zz}}
    \end{bmatrix}
    =
    \begin{bmatrix}
        \frac{\partial \mathrm{B_x}}{\partial x} & \frac{\partial \mathrm{B_x}}{\partial y} & \frac{\partial \mathrm{B_x}}{\partial z} \\[6pt] 
        \frac{\partial \mathrm{B_y}}{\partial x} & \frac{\partial \mathrm{B_y}}{\partial y} & \frac{\partial \mathrm{B_y}}{\partial z} \\[6pt] 
        \frac{\partial \mathrm{B_z}}{\partial x} & \frac{\partial \mathrm{B_z}}{\partial y} & \frac{\partial \mathrm{B_z}}{\partial z} \\[6pt] 
    \end{bmatrix}.
    \label{tensorEq}
\end{equation}

This technique, also called full tensor gradiometry, can provide benefits over vector measurements including improvements to spatial resolution \cite{gradiometry_spatail_res_improvments, magnetic_tensor_gradiometetry_small_scale_fields, foss_improvments_to_source_resolution}, directional filtering \cite{directional_filtering_gradiometry}, common mode noise rejection \cite{magnetic_gradiometry_common_mode_noise_rejection_example, magnetic_gradiometry_common_mode_noise_rejection_example_2} and decoupling the contributions from homogeneous background fields \cite{Schmidt2006TheCharacterization, Schmidt2004GETMAGExploration, Murphy2007TargetData}. Gradients due to local anomalies are larger than homogeneous background fields and therefore smaller features can be resolved where normally they would be dominated by the larger background.

Measuring the magnetic field components around an ensemble of NV centers in diamond can provide vector information instead of just the scalar projection along one of the four possible NV center axes in the diamond \cite{Schloss2018SimultaneousSpins}. NV vector magnetometry takes advantage of using the magnetic sensitivity of all four possible orientations of the defect rather than just a single orientation, increasing the probe population by four. Vector measurements are important in applications such as magnetic navigation \cite{Luo2015FullApplication}, battery monitoring \cite{battery_monitoring, battery_monitoring_2}, unexploded ordnance detection \cite{ordance_detection}, space missions, and geological surveys \cite{ Schmidt2004GETMAGExploration, Gang2015IntegratedSystem, Gang2016DetectionSystem, Young2010MagneticEnvironment} and could be applied to areas such as archaeology, non-destructive testing (NDT) \cite{Zhou2021ImagingMagnetometer}, magneto-cardiography (MCG) \cite{gavin_morley_mcg},  magneto-encephalography (MEG) \cite{Fagaly2006SuperconductingApplications, Korber2016SQUIDsHealthcare, AcunaMSpaceMag, Cochrane2016, Boto2018} and nuclear magnetic resonance spectroscopy \cite{Jiang2019MagneticResonance}. Magnetic imaging of damage in steel is of great interest to industry \cite{Hoult2014, Paulraj2013, Hu2022,Suresh2017DevelopmentTube,Kikuchi2010FeasibilityTube,Kikuchi2011FeasibilityPlants,Peng2020AnalysisAssessment,Liu2017TheMethod, Zhou2021ImagingMagnetometer}. 


Magnetic sensing is possible with NV centers in diamond through optically detected magnetic resonance (ODMR). Magnetometry with NV centers provides high dynamic range and the ability to perform measurements in a wide temperature range, in chemically harsh environments and under high irradiation. Ensembles of NV centers provide high magnetic sensitivities, with small sensing volumes e.g. $\mathrm{1~mm^3}$ or less \cite{Barry2020SensitivityMagnetometryb, Taylor2008High-sensitivityResolution, Acosta2009DiamondsApplications, Plakhotnik2014All-OpticalNanodiamond, Toyli2012MeasurementK, Toyli2013FluorescenceDiamond, Doherty2013TheDiamond, Hsieh2019ImagingSensor, Ivady2014PressureStudy, stuarts_sensitivity, Zhang_sensitivity}. NV center magnetometers have shown a range of applications including sensing induced eddy currents \cite{Chatzidrosos2019Eddy-CurrentDiamond, Ibrahim2018Room-TemperatureMagnetometry, Xu2022BurstMagnetometry}, single-neuron action potentials \cite{Barry2016OpticalDiamondb} and magnetic nanoparticle sensing in biomedical tissue \cite{Kuwahata2020MagnetometerApplications, Davis2018MappingMagnetometry}. 



Vector magnetometry with NV centers uses excitation of the ground state spin level transition $m_s = 0 \rightarrow m_s = \pm 1$ for at least three of the four possible orientations of the NV centers in diamond. This enables vector magnetometry with NV centers using a single sensor head, unlike other magnetometers such as fluxgates which require multiple sensors. This means that low non-linearity and non-orthogonality are built in. A majority of high-sensitivity NV center magnetometers are tabletop setups which are immobile preventing them from being used for scanning over a sample of interest.


Our fiber-coupled NV magnetometer setup can take vector measurements while scanning through three-dimensional space, rather than being fixed at a single point in space, allowing magnetic tensor gradiometry images to be recorded. A small magnet injects magnetic flux into the sample to be imaged, and distortions of this flux by the sample are imaged by the diamond magnetometer \cite{Zhou2021ImagingMagnetometer}.

\begin{figure}[t]
    \begin{subfigure}[t]{0.2\textwidth}
        \phantomsubcaption
        \label{fig:steelPlateDrawing} 
        \includegraphics[width = \textwidth]{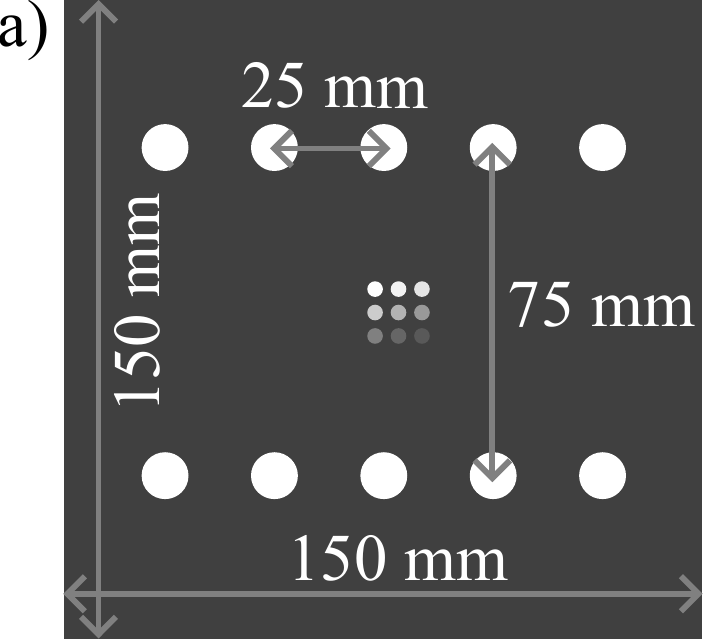}
    \end{subfigure}%
    
    \begin{subfigure}[t]{0.48\textwidth}
    \phantomsubcaption
    \label{fig:scanSetup} 
    \includegraphics[width = \textwidth]{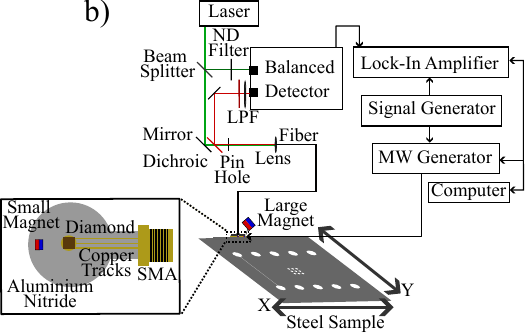}
    \end{subfigure}
    \caption{\textbf{a)} The 4.5~mm 316 steel plate used in this work. The large holes have a diameter of 6~mm and the small holes have a diameter of 2~mm with a separation of 10~mm. The small holes are drilled at varying depths from 4.5~mm (top left) to 0.5~mm (bottom right). \textbf{b)} Diagram of the scanning magnetometry setup (ND - neutral density, MW - microwave, LPF - low pass filter). The diamond sits on top of the microwave loop, which is printed onto an aluminum nitride base.}
\end{figure}

Here we present a scanning fiber-coupled NV magnetometer that is capable of both vector magnetometry and magnetic tensor gradiometry with a moving sensor head, independent of the main optoelectronics housed in a mobile equipment rack. Changes in the ODMR peak frequencies from magnetic field perturbations are tracked to calculate magnetic field vectors and gradients, while simultaneously moving the sensor head through the environment. As a demonstration of its capabilities, the magnetic vector measurements around a 316 stainless steel plate (Fig. \ref{fig:steelPlateDrawing}) are made. The plate has holes drilled in at various positions with a range of sizes to simulate damage. We use no magnetic shielding.


The fiber coupled scanning setup is shown in Fig. \ref{fig:scanSetup}. A Laser Quantum 532~nm GEM laser is used for excitation of the NV ensemble of the diamond, inside the sensor head. A laser power of 1~W is used with approximately 0.6~W of power measured at the diamond. The optoelectronics are kept on a mobile rack which allows for more portability. The fiber is a custom ordered FG910UEC 3~m fiber with a core diameter of 0.91~mm with steel ferrule-connector-physical-contact (FCPC) connectors on both ends. The fiber is directly coupled to the diamond on the sensor head via a SM1FC2 fiber adapter which is then screwed down onto the antenna substrate to make contact with the diamond. The fiber is secured to reduce modal noise from movement of the fiber during scanning of the sensor head. Fluorescence from the sensor head is sent back through the same fiber to be focused onto one of the two photodiodes on a Thorlabs PD450A balanced detector. The other photodiode takes a reference beam directly from the green laser which allows for cancellation of common mode noise, mainly laser noise. When the sensor head is on-resonance, the reference and fluorescence signals are balanced to get the best cancellation performance. The subtracted signal from the output of the balanced detector is then digitized and demodulated via a Zurich MFLI DC-500 kHz lock-in amplifier (LIA).

Microwaves are provided by an Agilent E8257D microwave source and amplified via a 43-dB Mini-Circuits ZHL-16W-43-S+. Microwaves are delivered to the diamond by a 2~m coaxial cable which connects to an SMA adapter soldered directly to a co-planar waveguide on an aluminum nitride ceramic printed circuit board. The waveguide then leads to a 3.0~mm diameter loop antenna with the diamond situated in the center. The aluminum nitride provides high thermal conductivity to act as a good heat sink, pulling heat from the diamond as it is heated by both the laser and microwaves. The diamond is a 1~mm cube. It is a low-strain diamond grown with chemical vapor deposition by Element Six, having (100) polished faces and 99.995\% $\mathrm{^{12}C}$ isotopic purification \cite{diamond_properties_edmonds_walsworth}. 

\begin{figure}[t]
    \centering
    \begin{subfigure}[t]{0.45\textwidth}
        \phantomsubcaption
        \label{fig:ODMROnOffSteel} 
        \includegraphics[width = \textwidth]{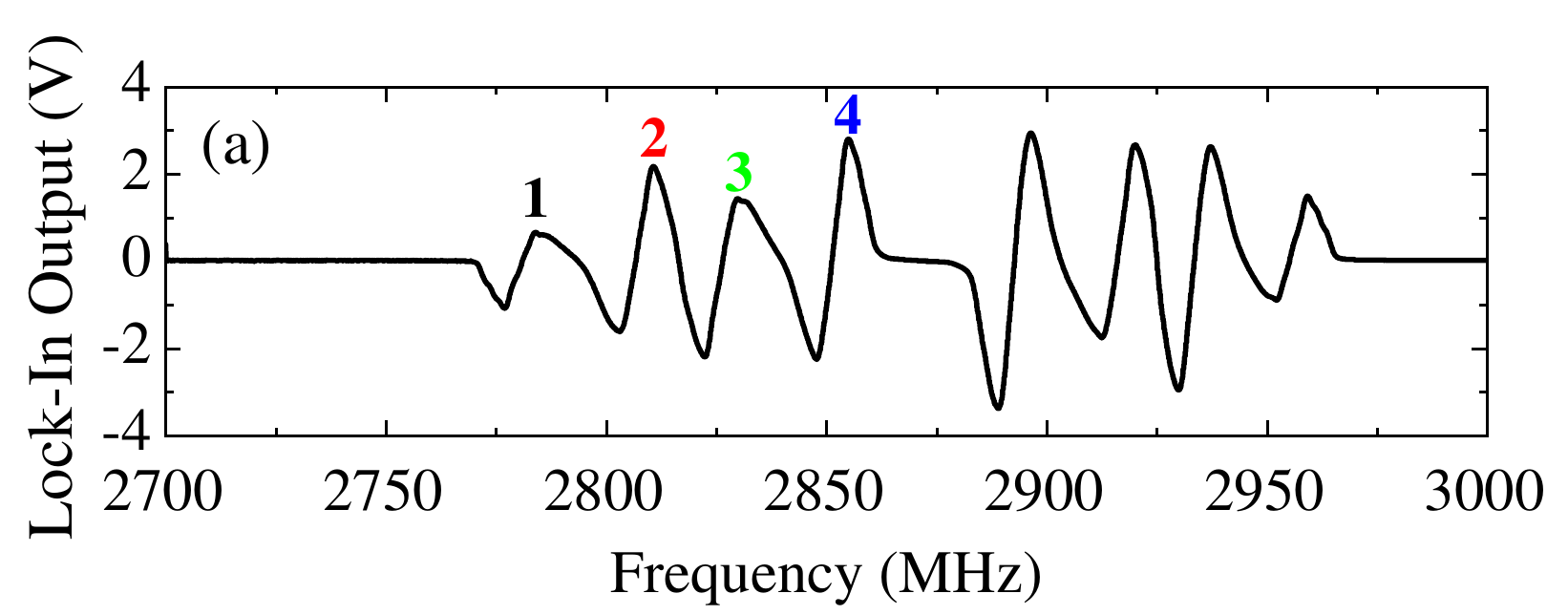}
    \end{subfigure}
    \begin{subfigure}[t]{0.45\textwidth}
        \phantomsubcaption
        \label{fig:4PeakTrackingFreqTimeTrace}
        \centering
        \includegraphics[width = 1\textwidth]{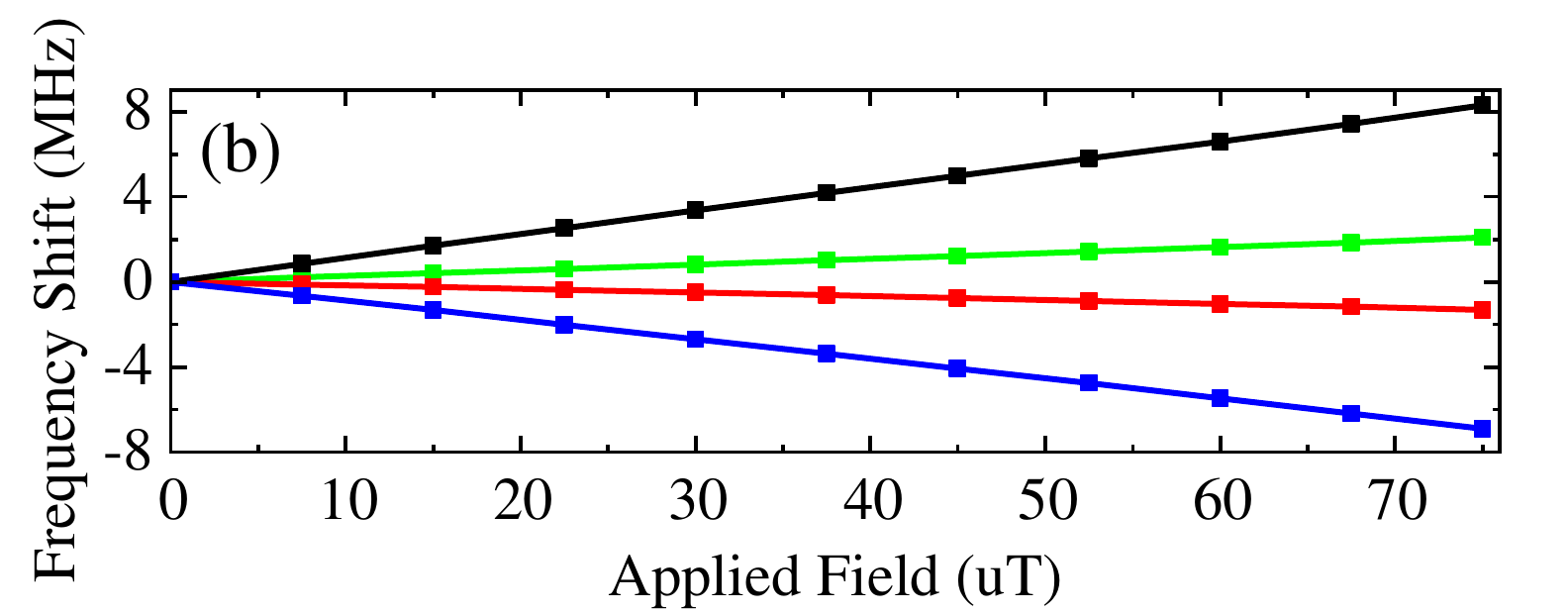}
    \end{subfigure}
    \caption{(a) Optically detected magnetic resonance spectrum used for vector measurements while the sensor head is on the steel. (b) Example of tracked frequency shifts of four peaks, as labeled in panel (a), as an applied solenoid field along the z axis is increased.}
\end{figure}

A permanent Nd-Fe-B magnet is attached to the side of the sensor head using a custom 3D printed mount which allows 360 degree rotation around the azimuthal angle and 180 degree rotation in the polar angle. The magnet is orientated until the bias field projection differs for all four crystallographic directions of the NV center, to provide all eight separate resonance peaks in the ODMR spectrum, as seen in Fig. \ref{fig:ODMROnOffSteel}. Parameter optimization was performed to achieve the best sensitivity for vector magnetometry, following the process outlined in \cite{Patel2020SubnanoteslaSensor}. With a low-pass filter (LPF) with a 150~Hz 3-dB point, the optimum parameters for microwave power after the amplifier, modulation frequency and modulation amplitude are found to be 30 dBm, 3.0 kHz and 3.0~MHz, respectively. The microwaves are sine wave frequency modulated. To determine the sensitivity during optimisation, a linear fit is made to the fourth ODMR peak (\ref{fig:ODMROnOffSteel}) and then fast Fourier transforms (FFT) of thirty 1 s time traces were taken and averaged, when on the central frequency of that peak. The mean sensitivity is taken as the average value between 10 and 150~Hz, excluding the 50~Hz peak due to the mains electricity.

A 3D printer stage is used to scan the sensor head in three axes. The stage allows for control over the step size and speed of the scan as well as the lift-off distance in the z axis. The step size used for the x and y movement was 0.5~mm.

\begin{figure*}[t]
    \begin{subfigure}[t]{0.32\textwidth}
        \phantomsubcaption     
        \label{fig:BxVectorImage}
        \includegraphics[width = \textwidth]{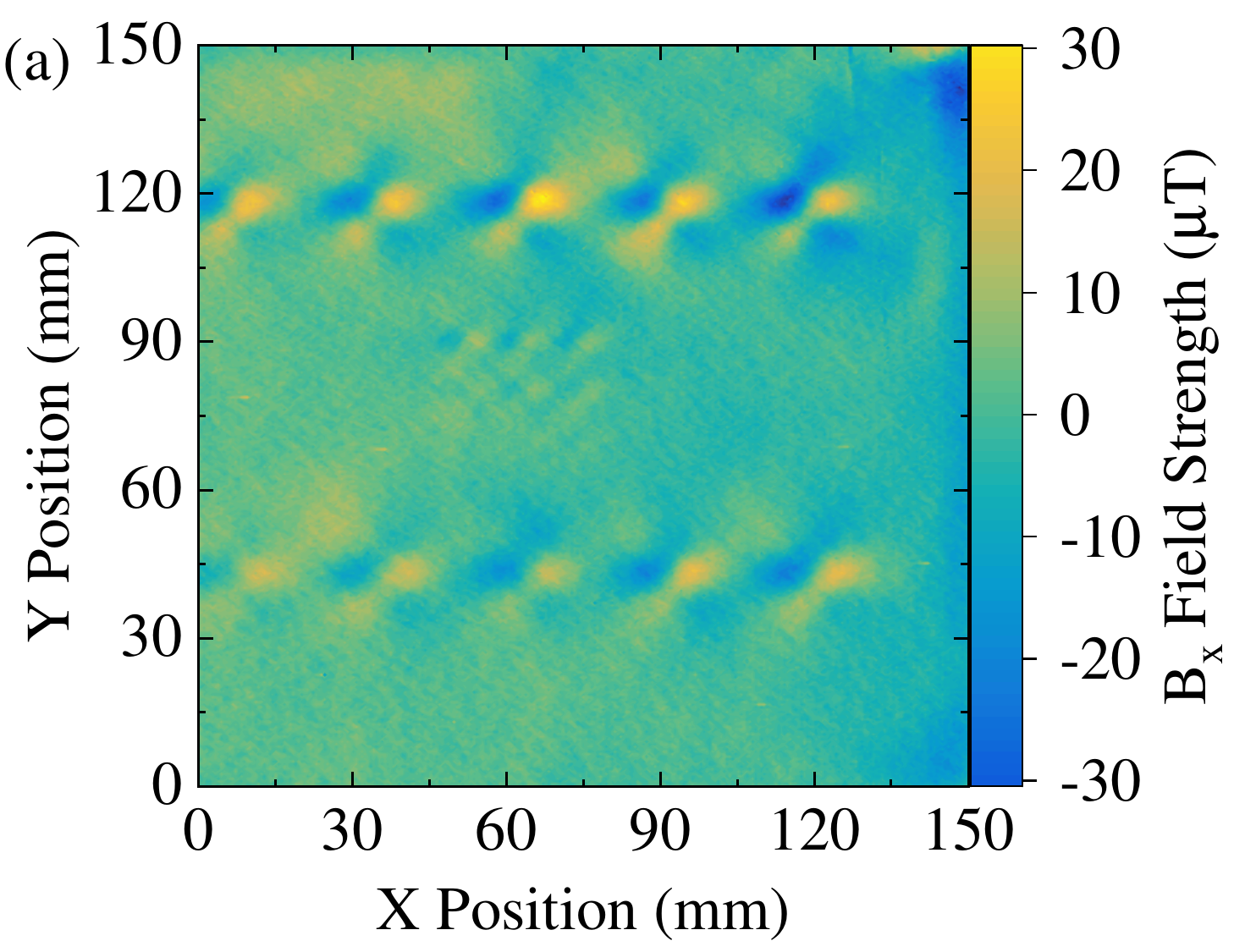}
    \end{subfigure}
    ~
    \begin{subfigure}[t]{0.32\textwidth}
        \phantomsubcaption
        \label{fig:HolePlateVectors_TMI} 
        \includegraphics[width = \textwidth]{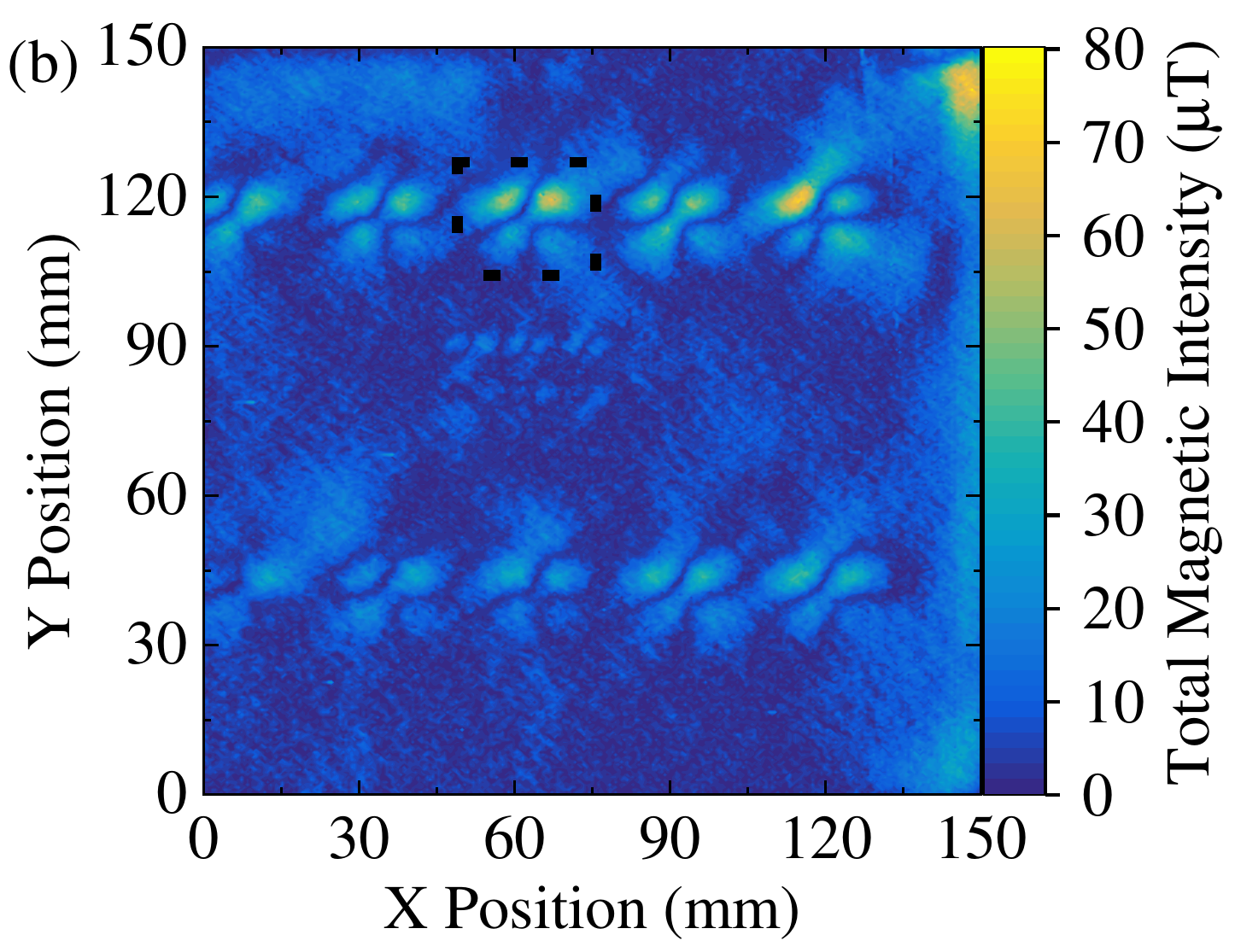}
    \end{subfigure}%
    ~
    \begin{subfigure}[t]{0.32\textwidth}
        \phantomsubcaption   
        \label{fig:HolePlateVectors_gradient}
        \includegraphics[width = \textwidth]{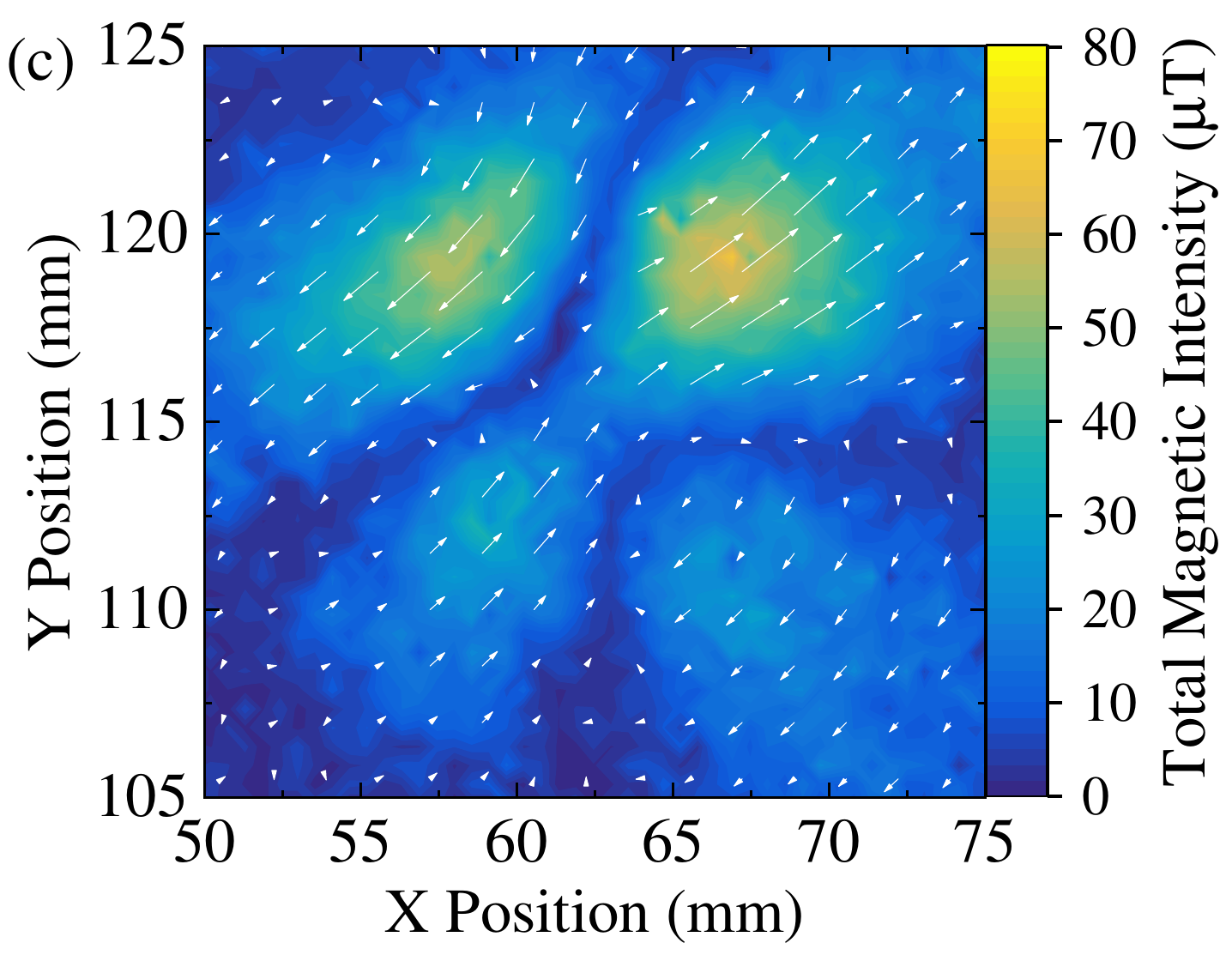}
    \end{subfigure}
    \caption{(a) Map of the $\mathrm{B_x}$ field strength. (b) Map of the total magnetic intensity. The highlighted box shows a region where a 6 mm hole is located. (c) Measured magnetic field vectors from the highlighted region. Collecting these images took 12 hours, which is nine times faster than in ref 21, due to reduced dwell time before moving to the next pixel.}
\end{figure*}

Changes to the external magnetic field are measured by looking at the change in peak resonance frequency due to Zeeman shift of the NV center $\mathrm{m_s = - 1}$ spin level. The shift in resonance frequency causes the fluorescence intensity to change which results in a change in the LIA voltage output. When scanning over a magnetic object, like the steel plate, the change in external field is caused by the difference in the amount of material near the sensor head. The changes in voltage are used to provide a process variable for a proportional linear feedback system which adjusts the microwave frequency to continually follow the central frequencies of the four chosen peaks. Using the peak's central resonance frequency as a set-point, any changes to the continually monitored voltage while at that central frequency can be used to determine the shift in resonance frequency, by using the known gradient of the ODMR peak. The calculated shift in resonance is used as an error signal which is sent to the microwave source to change the microwave frequency until the monitored voltage returns back to its set-point value, at the center of the peak. This is then expanded to cover four ODMR peaks by sequentially looping over each one, approximately every 70 ms, to measure their individual shifts, as shown in Fig. \ref{fig:4PeakTrackingFreqTimeTrace}. Tracking the four peaks can then be used to determine the magnetic field components along each of the four NV center axes which ultimately allows for calculation of the $\mathrm{B_x}$, $\mathrm{B_y}$, and $\mathrm{B_z}$ components of the field, relative to a coordinate frame defined around the sensor head. Another advantage to this resonance tracking technique is that it provides a much higher dynamic range of the magnetometer as the sensitive range is no longer limited to the linewidth, or linear region, of the ODMR peak. 

The vector components of an applied field are reconstructed using four ODMR line centers which each represent the four possible orientations of the NV center. The first four ODMR line centers used represent the $\mathrm{m_s = 0}$ to $\mathrm{m_s = -1}$ transitions of the four orientations (Fig. \ref{fig:ODMROnOffSteel}). The method to calculate the vector components was taken from the work done by Schloss et al. \cite{Schloss2018SimultaneousSpins}. The linearized NV center ground-state Hamiltonian allows magnetic field components to be determined from the frequency shifts of the four peaks. The result is

\begin{equation}
    \begin{bmatrix}
     \mathrm{B_x} \\
     \mathrm{B_y} \\
     \mathrm{B_z} \\
    \end{bmatrix}
    = \mathbf{A^{-1}} 
    \begin{bmatrix}
    \Delta f_1 \\
    \Delta f_2 \\
    \Delta f_3 \\
    \Delta f_4 \\
    \end{bmatrix}
    ,
    ~\mathbf{A} =
    \begin{bmatrix}
        \Large
        \frac{\partial f_1}{\partial \mathrm{B_x}} & \frac{\partial f_1}{\partial \mathrm{B_y}} & \frac{\partial f_1}{\partial \mathrm{B_z}} \\[6pt]
        \frac{\partial f_2}{\partial \mathrm{B_x}} & \frac{\partial f_2}{\partial \mathrm{B_y}} & \frac{\partial f_2}{\partial \mathrm{B_z}} \\[6pt]
        \frac{\partial f_3}{\partial \mathrm{B_x}} & \frac{\partial f_3}{\partial \mathrm{B_y}} & \frac{\partial f_3}{\partial \mathrm{B_z}} \\[6pt]
        \frac{\partial f_4}{\partial \mathrm{B_x}} & \frac{\partial f_4}{\partial \mathrm{B_y}} & \frac{\partial f_4}{\partial \mathrm{B_z}}
    \end{bmatrix}
    \label{vecCalc}
    ,
\end{equation}

where $\Delta f_i$ is the shift in frequency of peak $i = 1,2,3,4$ and $\mathbf{A^{-1}}$ is the psuedo-inverse of matrix $\mathbf{A}$ where $\mathrm{B_x}$, $\mathrm{B_y}$ and $\mathrm{B_z}$ are the magnetic field components in the frame defined around the sensor head.

To calculate matrix $\mathbf{A}$, first the sensor head was placed inside a Helmholtz coil with a calibration of 78.2~$\mathrm{\mu T / A}$ between 0.1 and 1~A. The calibration was performed with a Bartington Instruments Mag-03MS100 three-axis fluxgate magnetometer. The coil was aligned along the defined x axis relative to the sensor head and the peak shifts were tracked as the current was increased in steps of 0.1~A. This was repeated for the y and z axes. A linear fit for all four peaks for each axis orientation was made to determine the change in peak central frequency with the increase in applied field, as shown in Fig. \ref{fig:4PeakTrackingFreqTimeTrace}. The matrix $\mathbf{A}$, from the twelve fits, was found to be

\begin{equation}
    \mathbf{A} \, (\mathrm{MHz/uT}) =
    \begin{bmatrix}
         0.12 & 0.04 & -0.13 \\
         -0.02 & -0.10 & -0.09 \\
         0.03 & 0.11 & -0.19 \\
         -0.09 & -0.03 & -0.14
    \end{bmatrix} 
    .
\end{equation}


With this calibration matrix it is then possible to track the peak shifts and calculate the vector components of the field which caused said shift. We do this in real-time.


A 160~mm x 160~mm scan was done across the 150~mm x 150~mm 316 stainless steel plate, with a step size of 0.5~mm for both the x and y directions and with a separation distance of approximately 1~mm between the surface of the steel sample and the base of the antenna. The antenna is 1~mm thick so the sensing diamond is 2~mm away from the sample. A total magnetic intensity (TMI) map is produced by using $\mathrm{|B| = \sqrt{B_{x}^2 + B_{y}^2 + B_{z}^2}}$. 

Both the large and small holes are successfully imaged in all three vector components; an example of $\mathrm{B_x}$ is shown in Fig. \ref{fig:BxVectorImage}. The features are also clearly visible in the TMI map, Fig. \ref{fig:HolePlateVectors_TMI}, where the change in field magnitude indicates a lack of steel. With the addition of vector magnetometry we are also able to plot the magnetic field vectors which can help with locating and analyzing defects, as shown in Fig. \ref{fig:HolePlateVectors_gradient}. The magnetic field preferentially flows through the higher permeability steel, as opposed to the air and this can be visualized with the field vectors as they move away and around the hole feature in the map to stay flowing through the steel. This can be useful for detecting defects not directly visible by giving information on whether the features seen in a magnetic field map are defects due to loss of material or are just regions of lower magnetic field intensity, where you would see the field vectors pass unperturbed.

\begin{figure}
    \begin{subfigure}[t]{0.23\textwidth}
        \phantomsubcaption
        \label{fig:bx_newDataZoomed}
        \includegraphics[width = \textwidth]{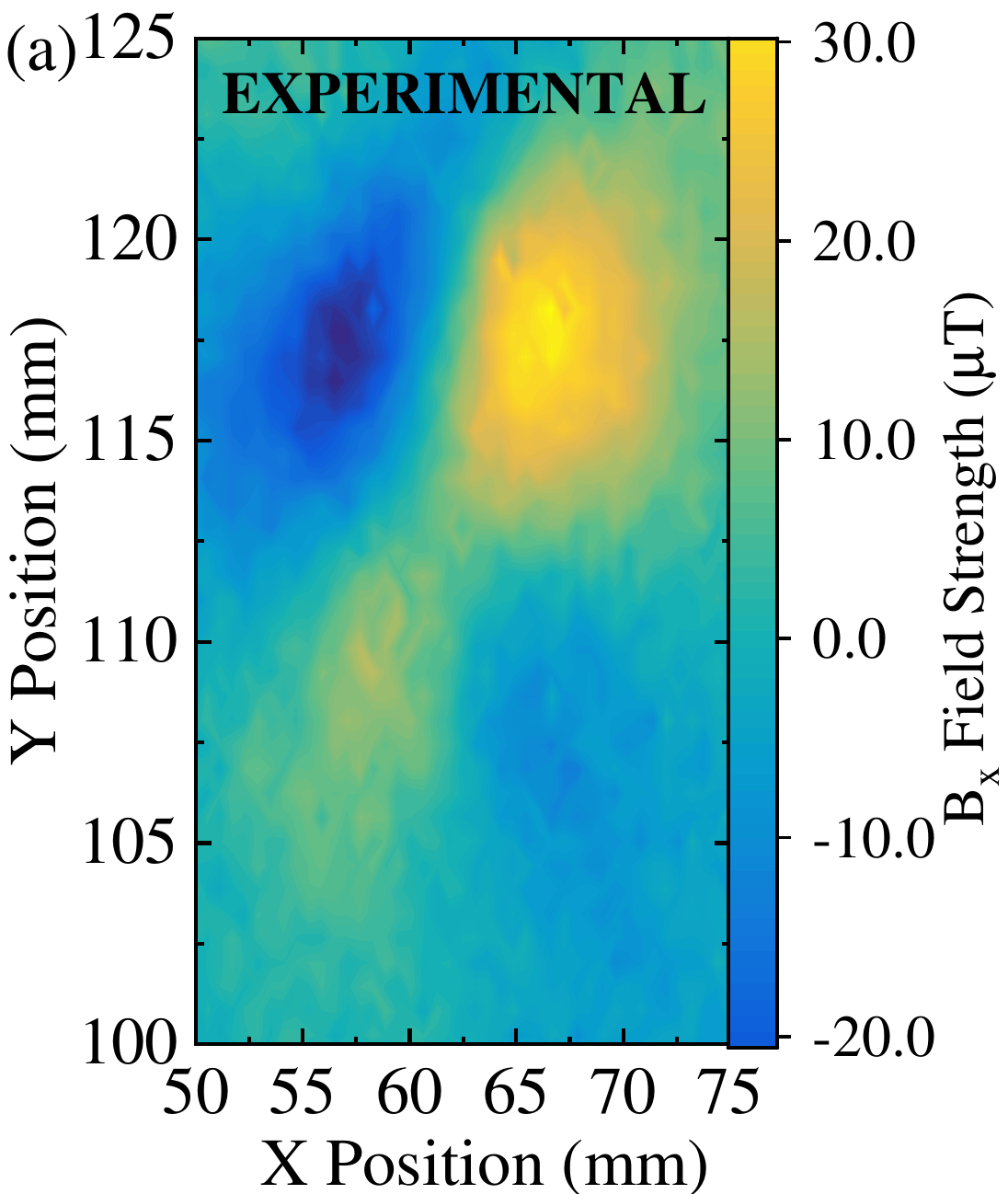}
    \end{subfigure}
    ~
    \begin{subfigure}[t]{0.23\textwidth}
        \phantomsubcaption
        \label{fig:bx_oldDataZoomed}
        \includegraphics[width = \textwidth]{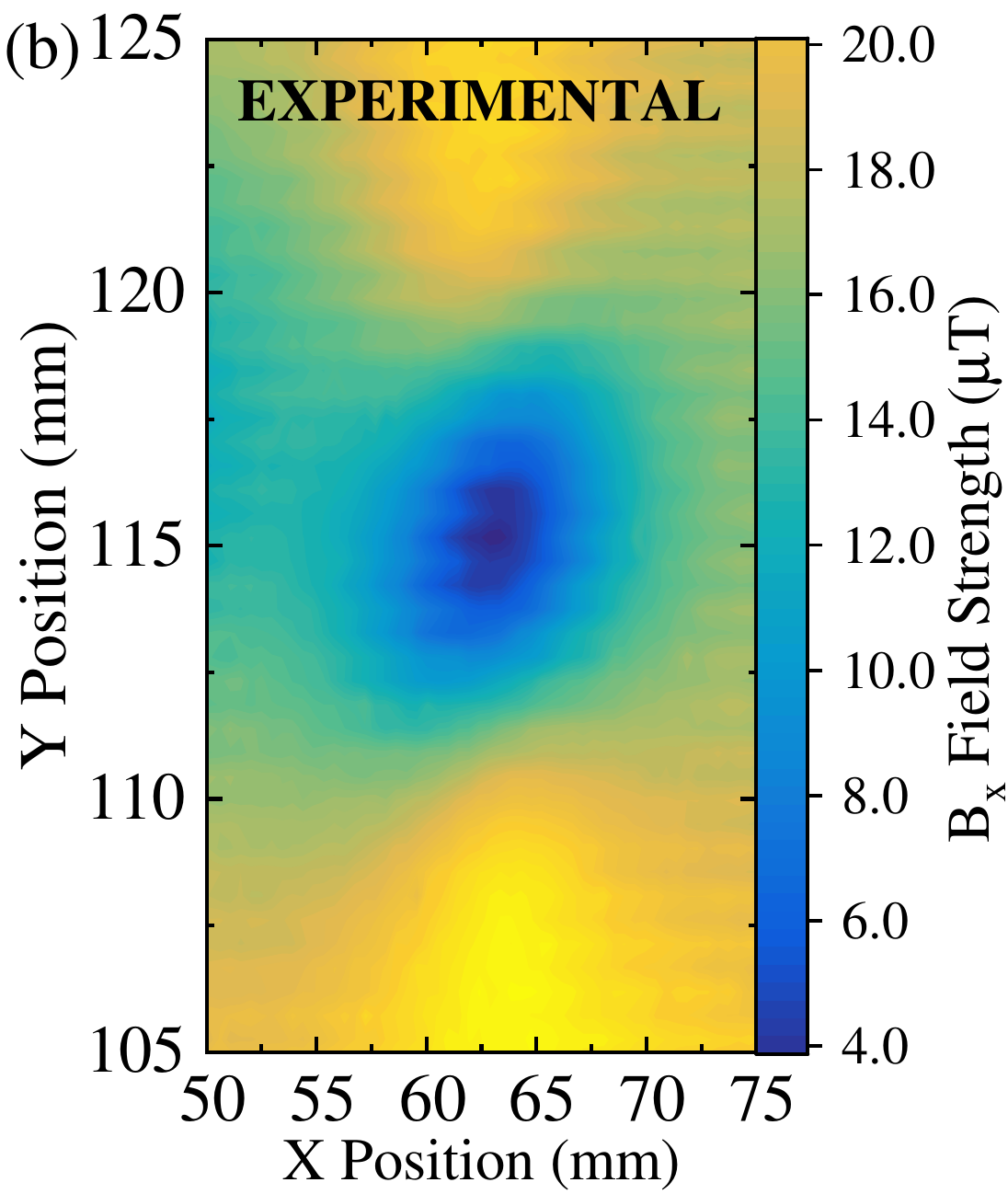}
    \end{subfigure}

    \begin{subfigure}[t]{0.23\textwidth}
        \phantomsubcaption
        \label{fig:bx_comsol_new}
        \includegraphics[width = \textwidth]{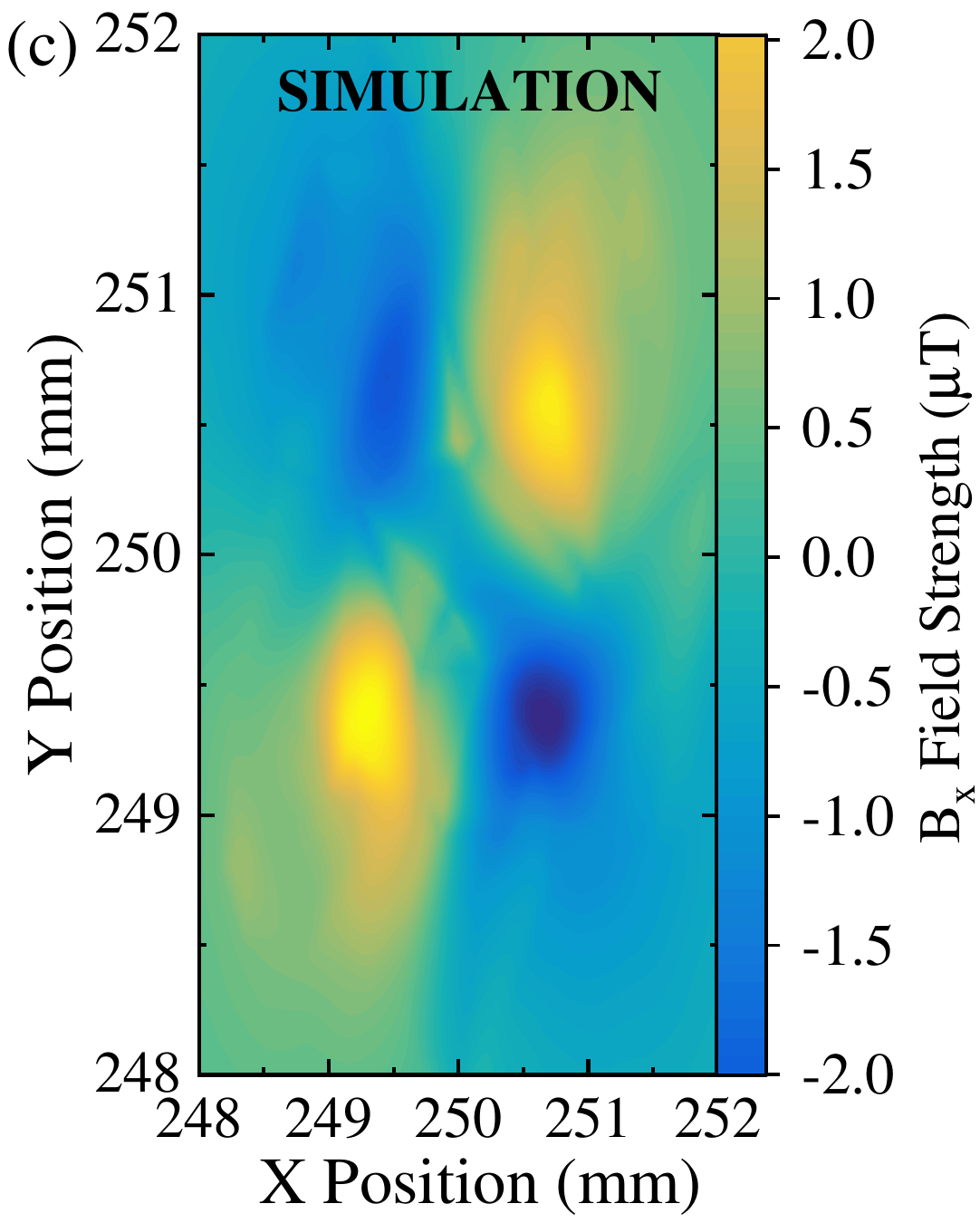}
    \end{subfigure}
    ~
    \begin{subfigure}[t]{0.23\textwidth}
        \phantomsubcaption
        \label{fig:bx_comsol_old}
        \includegraphics[width = \textwidth]{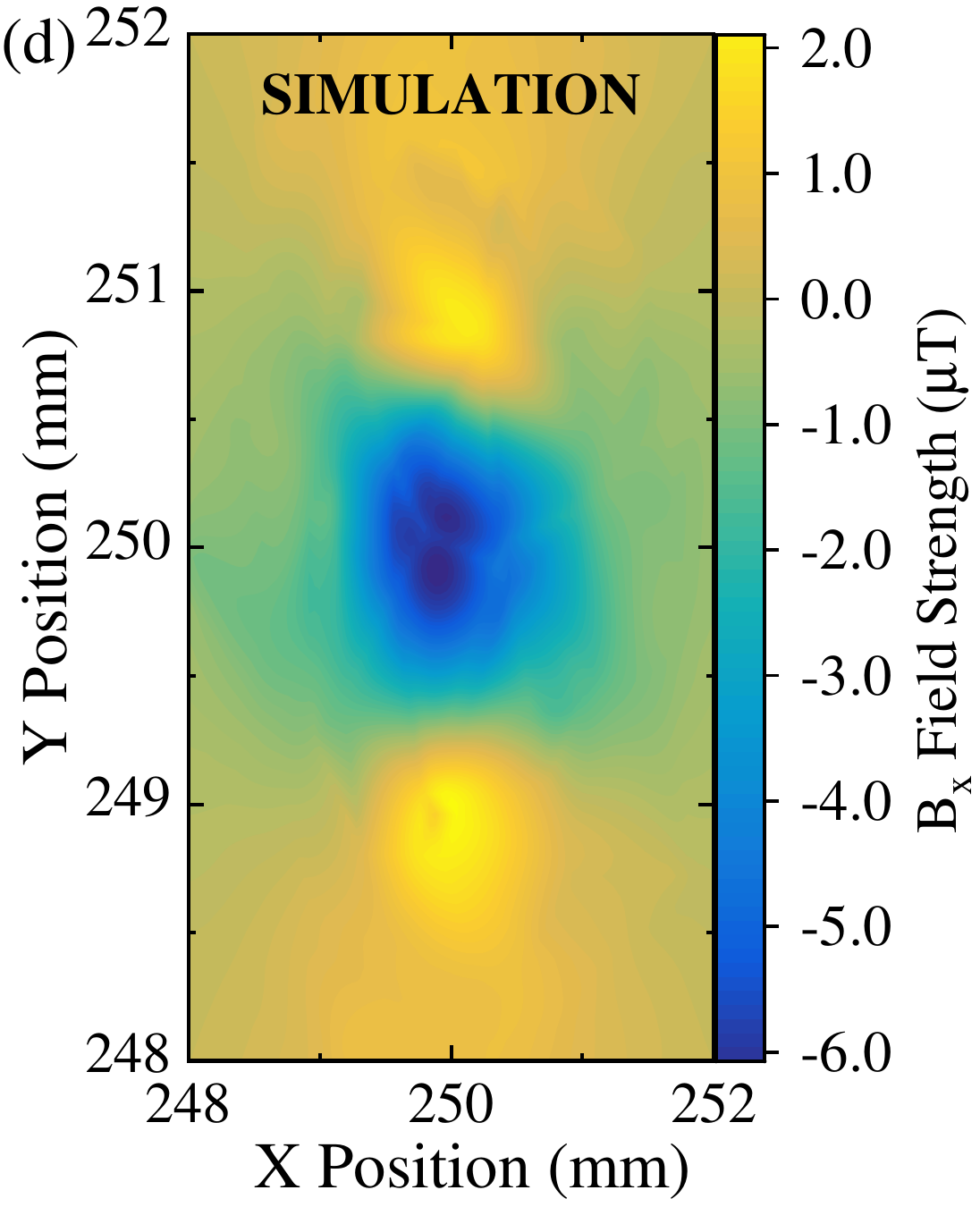}
    \end{subfigure}
    
    \caption{(a) $\mathrm{B_{x}}$ field component image over a single 6~mm diameter hole for the unrotated small bias magnet, showing quadrupole pattern. (b)  $\mathrm{B_{x}}$ field component image over a single large hole for the rotated small bias magnet, showing dipole pattern. (c) COMSOL simulation of the $\mathrm{B_{x}}$ field over a large hole with an unrotated magnet. (d) COMSOL simulation of the $\mathrm{B_{x}}$ field over a large hole with a rotated magnet.}
    \label{holePlate_TensorMaps}
\end{figure}

Two horizontal lines of evenly spaced quadrupole shapes at the top and bottom of the plate indicate the locations of the large holes in the steel, with the grid of nine holes in the center. The quadrupole field shape coming from the holes (Fig. \ref{fig:HolePlateVectors_gradient} and Fig. \ref{fig:bx_newDataZoomed}) is found to be due to the orientation of the small bias magnet relative to the hole features. Rotating the magnet 45 degrees, between having its square face facing the diamond and having it edge-on to the diamond, produces different field component and TMI images, as seen in Fig. \ref{fig:bx_oldDataZoomed}, where the $\mathrm{B_x}$ field component now has a more dipole-like shape. By performing COMSOL simulations, it was verified that the rotation of the small bias magnet was the cause of the change in field pattern seen. Simulations of a 316 stainless steel plate with a 6~mm hole and a 1~mm cube 0.1~T permanent magnet placed 1~mm above the hole matches well with what is measured when the magnet is not rotated (Fig. \ref{fig:bx_comsol_new}) and with a rotation of 45 degrees (Fig. \ref{fig:bx_comsol_old}).



\begin{figure}
    \begin{subfigure}[t]{0.36\textwidth}
        \phantomsubcaption
        \label{fig:bxy}
        \includegraphics[width = \textwidth]{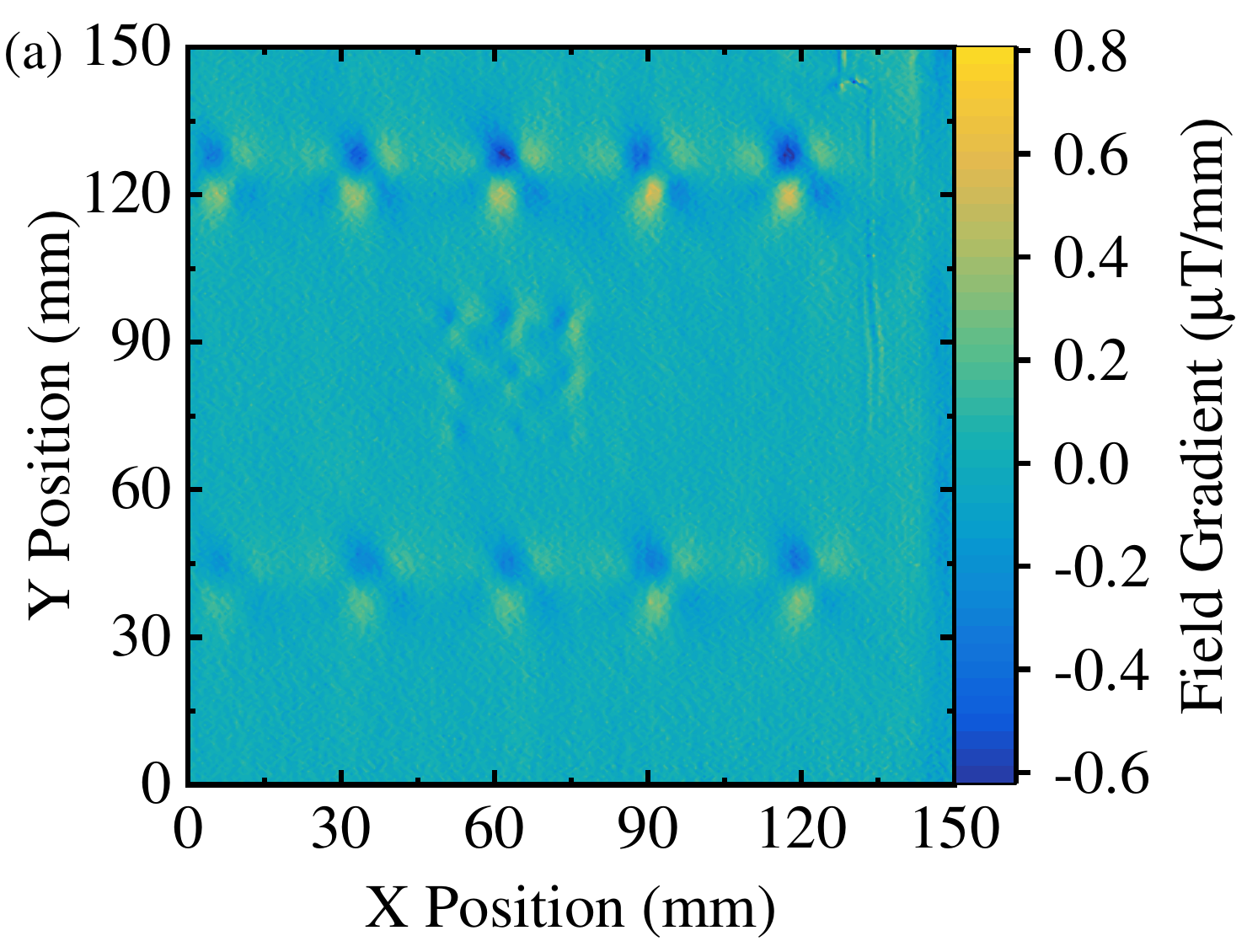}
    \end{subfigure}
    
    \begin{subfigure}[t]{0.46\textwidth}
        \phantomsubcaption
        \label{fig:VLineSmall}
        \includegraphics[width = \textwidth]{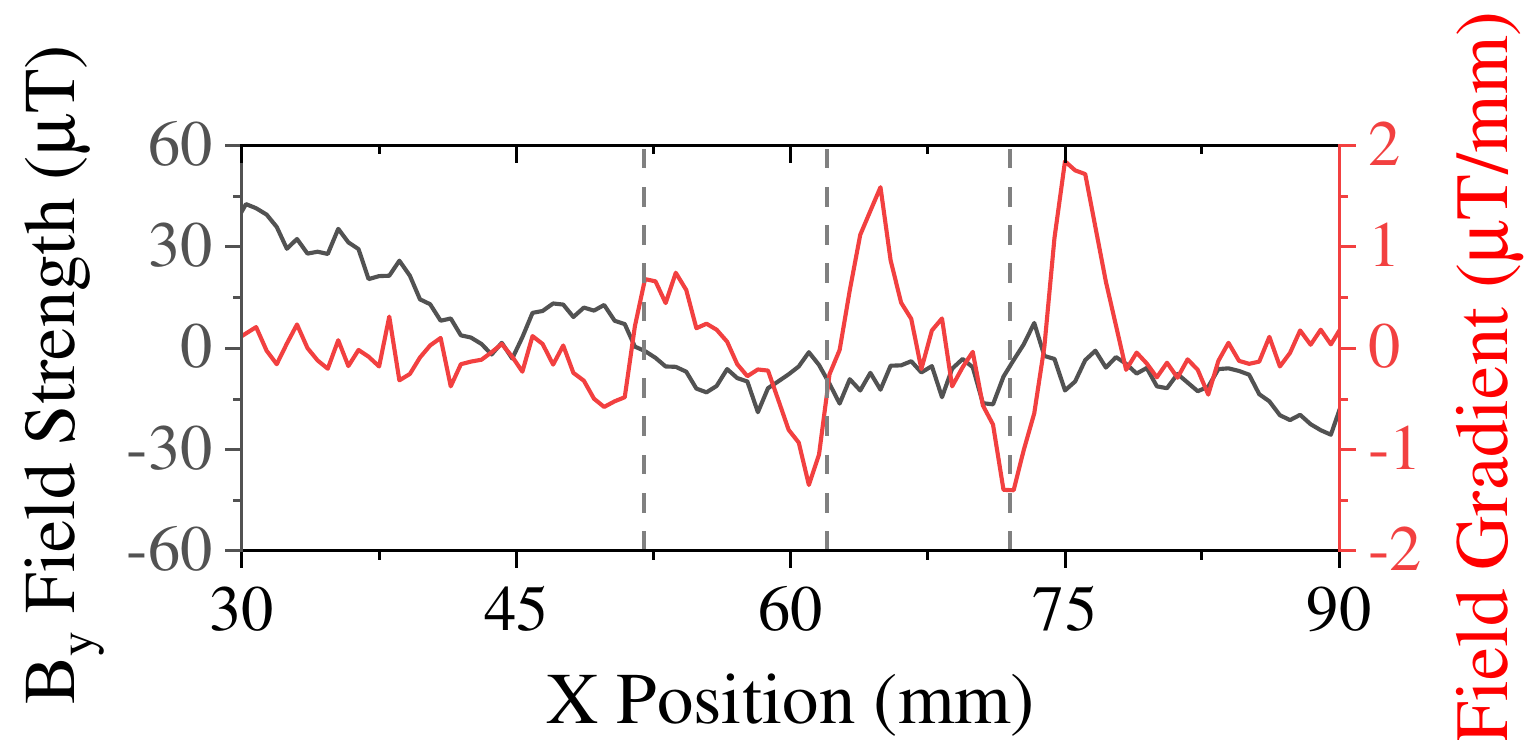}
    \end{subfigure}
    \caption{\textbf{a)} $\mathrm{B_{xy}}$ tensor component map. \textbf{b)} A $\mathrm{y =~97}$~mm horizontal line profile over top row of central small 2~mm diameter holes, for both the $\mathrm{B_y}$ (black), the $\mathrm{B_{xy}}$ component (red).}

\end{figure}

Scanning vector magnetometry makes it possible to perform tensor gradiometry. We have three matrices each containing the $\mathrm{B_x}$, $\mathrm{B_y}$ and $\mathrm{B_z}$ field components at different adjacent x, y and z positions so we can take the gradient of the field components with respect to the three Cartesian directions and generate nine tensor components for each point in space.


    

An example of this can be seen in Fig. \ref{fig:bxy} where the $\mathrm{B_{xy}}$ tensor component from Eq. \ref{tensorEq} has been plotted and it is clear that some features, for example the three lowest holes in the smaller grid of nine 2~mm holes in the center of the plate, are shown more clearly when compared to the TMI map (Fig. \ref{fig:HolePlateVectors_TMI}) and $\mathrm{B_{x}}$ map (Fig. \ref{fig:BxVectorImage}) where the three lowest of the smaller central holes are virtually impossible to resolve. The tensor components for the change in z direction were also calculated but did not provide any useful information, for more on this see the appendix. To compare this with the $\mathrm{B_{x}}$ map, a horizontal line profile on the $B_{xy}$ map (Fig. \ref{fig:bxy}) is taken across the small holes at the same y positions as on the  $\mathrm{B_{x}}$ map.

In conclusion, magnetic tensor gradiometry was shown to improve the imaging of small holes in steel, making them easier to characterize.

The authors thank Jeanette Chattaway, Lance Fawcett and Matty Mills of the Warwick Physics mechanical workshop. We are grateful for insightful discussions with Rajesh Patel throughout this work. Alex Newman's Ph.D studentship is funded by an EPSRC iCASE award to NNL (the National Nuclear Laboratory). Stuart Graham's Ph.D studentship is funded by DSTL (the Defence Science and Technology Laboratory).  This work is supported by the UK Networked Quantum Information Technologies (NQIT) Hub, and the UK Hub in Quantum Computing and Simulation, part of the UK National Quantum Technologies Programme, with funding from UKRI EPSRC grants EP/M013243/1 and No. EP/T001062/1 respectively. This work is also supported by Innovate UK grant 10003146, EPSRC grant EP/V056778/1 and an EPSRC Impact Acceleration Account (IAA) award. GWM is supported by the Royal Society.

\end{document}